\documentclass[prl,aps,amsmath,amssymb,amsfonts,showpacs,superscriptaddress,
twocolumn]{revtex4}
\input{epsf}
\usepackage{psfrag,graphicx}
\usepackage{epsfig}
\newcommand{\sub}[1]{\mbox{\tiny{#1}}}
\def\bavs3{BaVS$_3$}
\def\Tmit{$T_{\sub{MIT}}\,$}
\def\vk{{\bf k}}
\def\a1g{$A_{1g}$}

\begin{document}

\title{Importance of interorbital charge transfers for the metal-to-insulator
transition of BaVS$_3$}
\author{Frank Lechermann}
\affiliation{CPHT \'Ecole Polytechnique, 91128 Palaiseau Cedex, France}
\affiliation{LPT-ENS, 24 Rue Lhomond, 75231 Paris Cedex 05, France}
\author{Silke Biermann}
\affiliation{CPHT \'Ecole Polytechnique, 91128 Palaiseau Cedex, France}
\affiliation{Laboratoire de Physique des Solides, CNRS-UMR 8502, UPS
B\^{a}timent 510, 91405 Orsay, France}
\author{Antoine Georges}
\affiliation{CPHT \'Ecole Polytechnique, 91128 Palaiseau Cedex, France}
\affiliation{Laboratoire de Physique des Solides, CNRS-UMR 8502, UPS
B\^{a}timent 510, 91405 Orsay, France}
\begin{abstract}
The underlying mechanism of the metal-to-insulator transition (MIT) in BaVS$_3$
is investigated, using dynamical mean-field theory in combination with 
density functional theory. It is shown that correlation effects  
are responsible for a strong charge redistribution, which lowers the occupancy 
of the broader \a1g band in favor of the narrower $E_g$ bands. This resolves 
several discrepancies between band theory and the experimental findings,
such as the observed value of the charge-density wave ordering vector 
associated with the MIT, and the presence of local moments in the metallic 
phase.
\end{abstract}

\pacs{71.30.+h, 71.15.Mb, 71.10.Fd, 75.30.Cr}
\maketitle
%
%
The structural, electronic and magnetic properties of the vanadium
sulphide compound \bavs3 raise several puzzling
questions~\cite{whangbo_bavs3_puzzling_jsschem_2003,booth_bavs3_field_prb_1999}.
At room temperature, this material crystallizes in a hexagonal
($P6_3/mmc$) structure~\cite{gardner_actacry_1969}, in which straight chains of
face-sharing VS$_6$ octahedra are directed along the $c$-axis. At
$T_{\sub{S}}$$\sim$240 K the crystal structure transforms into an
orthorhombic ($Cmc2_1$) structure~\cite{ghedira_bavs3_neutrons_jpc_1986},
thereby creating an anisotropy in the $ab$-plane (i.e., perpendicular to the
chain direction) and a zigzag distortion of the VS$_3$ chains in the $bc$-plane.
Additionally, the Hall coefficient changes sign from negative to positve at
$T_{\sub{S}}$ \cite{booth_bavs3_field_prb_1999}. On further cooling the system
displays a metal-to-insulator transition (MIT) at \Tmit$\sim$$70$~K.
Remarkably, this transition is second-order and is not accompanied by magnetic
ordering. Only below $T_{\sub{X}}$$\sim 30$~K indications for an
incommensurate antiferromagnetic order exist~\cite{nakamura_bavs3_jpsj_2000}.

%
%
Forr\'{o} {\it et al.}~\cite{forro_bavs3_qcp_prl_2000} have found that the MIT
can be driven to $T$=$0$ by applying
pressure~\cite{graf_bavs3_pressure_prb_1995}. Further recent
experiments~\cite{inami_bavs3_prb_2002,fagot_bavs3_prl_2003} have
demonstrated that the MIT is in fact associated with a structural transition.
These studies establish that a commensurate structural modulation sets in,
corresponding to a reduced wave vector ${\bf q}$=(1,0,$\frac{1}{2})_O$ in
the orthorhombic cell. Furthermore, the X-ray diffuse scattering experiments of
Fagot et {\sl al.}~\cite{fagot_bavs3_prl_2003} reveal a large fluctuation regime
with critical wave vector $q_c$=$0.5c^*$(here $c^*$ is
the reciprocal unit vector along the $c$-axis of the orthorhombic system),
extending up to $170$~K into the metallic phase. In the same temperature range
the Hall coefficient is strongly increasing \cite{booth_bavs3_field_prb_1999}.
This regime might be interpreted as a precursor of the charge density wave
(CDW) instability, reminiscent of the large fluctuations in a quasi
one-dimensional (1D) metal in the vicinity of a Peierls transition. It should
be kept in mind however that the conduction anisotropy within the system is not
strongly pronounced
($\sigma_c/\sigma_a$$\sim$3-4)~\cite{mihaly_bavs3_prb_2000}, making 1D
interpretations questionable. The ``metallic'' phase above \Tmit displays
several other unusual
properties~\cite{graf_bavs3_pressure_prb_1995,mihaly_bavs3_prb_2000}.
The resistivity is rather high (a few m$\Omega$cm) and metallic-like 
($d\rho/dT$$>$$0$) only above $\sim$$150$~K. It displays a weak
minimum at this temperature, below which it increases upon further cooling.
Most interestingly, this phase displays
local moments, as revealed by the Curie-Weiss form of the magnetic
susceptibility. The effective moment corresponds approximately to one
localized spin-$1/2$ per two V sites. Since the formal valence is
V$^{4+}$, corresponding to one electron in the 3$d$-shell,
this can be interpreted as the effective localization of half of the electrons.
At \Tmit, the susceptibility rapidly drops, and the electronic entropy is 
strongly suppressed~\cite{imai_bavs3_entropy_jpn_1996}.

In the hexagonal high-$T$ phase the low-lying V(3$d$) levels consist of an
\a1g state and two degenerate $E_g$ states. A further splitting of the
degenerate states occurs in the orthorhombic phase. First-principles
calculations of the electronic structure of \bavs3, based on density functional
theory (DFT) in the local (spin) density approximation (L(S)DA), have been
performed in Refs.~\cite{nak94,mattheiss_bavs3_1995,whangbo_bavs3_jsschem_2002}.
For both phases, the calculations do yield a V(3$d$)-S(3$p$) hybridization
which is strong enough to account for the weak anisotropy of the transport
properties. No band-gap opening has been reached within L(S)DA. Instead, very
narrow $E_g$ bands right at the Fermi level, and a nearly filled dispersive
band with mainly \a1g character extending along the $c^*$ direction have been
found. This is consistent with a simple model proposed early on by
Massenet {\it et al.}~\cite{massenet_jpcsol_1979}. However, the
occupancy of the narrow $E_g$ bands found within LDA is too low to
account for the observed local moment in the metallic phase.
The nature of the CDW instability is also left unexplained
by the DFT calculations. Indeed, the
calculated value of the Fermi wave vector of the broad \a1g band is
found to be $2k_F^{\sub{LDA}}$$\simeq$$0.94c^*$~\cite{mattheiss_bavs3_1995},
while the observed wave vector of the instability is
$q_c$=$0.5c^*$~\cite{fagot_bavs3_prl_2003}. Therefore, the simple
picture of a CDW at $q_c$=$2k_F$ associated only with the \a1g band
is untenable within LDA. It is likely that the $E_g$ states also participate in
the instability, but the LDA band structure does not provide a Fermi-surface 
nesting compatible with the
experimental $q$ value~\cite{whangbo_bavs3_puzzling_jsschem_2003}.
Hence, ab-initio calculations based on L(S)DA are not sufficient to explain the
electronic structure of \bavs3. By using static DFT+U schemes a band gap was 
obtained~\cite{jia04}. However, this required to enforce magnetic order, hence 
leaving the question of the mechanism of the transition into the paramagnetic 
insulator unanswered~\footnote{Ref.~\cite{jia04} reported that without magnetic
order, the filling of the $A_{1g}$ band is actually further {\it increased}, 
in comparison to LDA.}.

In this letter, we present calculations in the framework of dynamical mean field 
theory (DMFT), using the LDA electronic structure as a starting point. On 
the basis of this LDA$+$DMFT treatment we propose correlation effects in a 
multi-orbital context as an explanation for the discrepancies between 
band theory predictions and experiments. Specifically, we show that
interorbital charge transfers occur, driven by the on-site Hund's coupling.
This lowers the occupancy of the \a1g orbital in favor of the $E_g$'s, hence
shifting $k_F(A_{1g})$ towards lower values. From a calculation of the local
susceptibilities, we demonstrate that local moments are formed in the metallic
phase, due to the low quasiparticle coherence scale induced by the strong
correlations (particularly for the narrow $E_g$ bands).

\begin{figure}
\epsfclipon
\epsfig{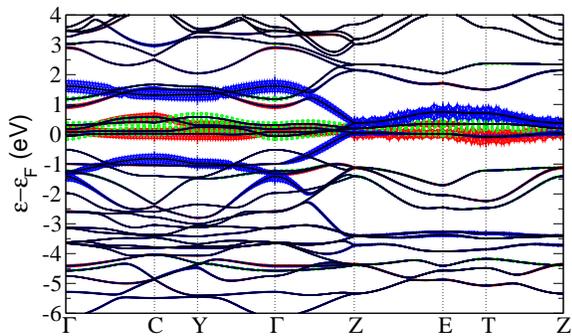}
\caption{ LDA band structure for BaVS$_3$ in the $Cmc2_1$ structure. Also shown
are the fatbands (see text) for the $A_{1g}$ (blue/full dark), $E_{g1}$
(red/full grey) and $E_{g2}$ (green/dashed grey) orbital of the V
atoms.}
\label{fig:fatbands}
\end{figure}

%
%
On Fig.~\ref{fig:fatbands} the band structure of \bavs3 calculated within LDA
for the $T$=100~K orthorhombic ($Cmc2_1$) crystal
structure~\cite{ghedira_bavs3_neutrons_jpc_1986} is displayed.
%
\begin{figure}
\epsfclipon
\epsfig{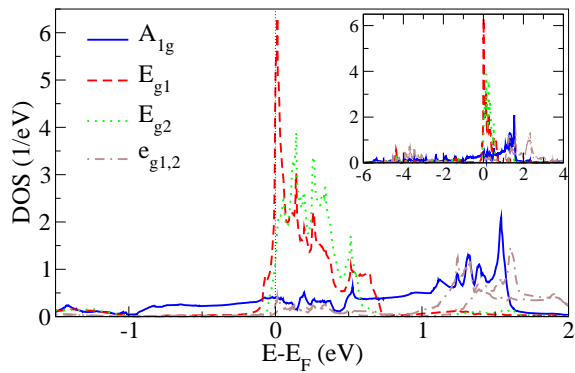}
\caption{ Angular momentum-resolved LDA-DOS of the symmetry-adapted V($3d$)
         states.}
\label{fig:vdos}
\end{figure}
The calculations were performed by using norm-conserving pseudopotentials
and a mixed basis consisting of plane waves and localized
functions~\cite{mbpp_code}. The results are consistent with previous
work~\cite{nak94,mattheiss_bavs3_1995,whangbo_bavs3_jsschem_2002}. The
symmetry-adapted V(3$d$)-basis $\{\phi_m\}$ was obtained by diagonalising the
orbital density matrix $n_{MM'}$$\sim$
$\sum_{{\bf k}b}f_{\bf k}^b\langle\psi_{\bf k}^b|M\rangle\langle M'|
\psi_{\bf k}^b\rangle$, where  $\psi_{\bf k}^b$ stands for the pseudo crystal
wave function for wave vector ${\bf k}$ and band $b$, and $M$,$M'$ denote the
cubic harmonics for $l$=2.
Being directed along the chain direction between pairs of V atoms, the
\a1g orbital has mainly $d_{z^2}$ character. In contrast, the $E_g$ states,
linear combinations of $d_{yz},\,d_{x^2-y^2}$ and $d_{z^2}$ ($E_{g1}$) as well
as $d_{xy}$ and $d_{xz}$ ($E_{g2}$), only weakly hybridize with their
surrounding.
The orbitals of the remaining $e_g$ manifold point mainly towards the sulphur
atoms, which results in a large energy splitting, leading to a smaller(larger)
contribution to the occupied(unoccupied) states well below(above) 
$\varepsilon_F$. Hence, the $e_g$ states do not have
a major influence on the essential physics around the MIT.
Note that in the orthorhombic structure (as already in the hexagonal one), the
unit-cell contains two formula units, with the two V-sites equivalent by
symmetry. The high-symmetry points $\Gamma$-C-Y in the Brillouin Zone (BZ)
define a triangle in the $k_z$=0 plane, whereas Z-E-T is the analogous shifted
triangle in the $k_z$=$0.5c^*$ plane. The $\Gamma$-Z line corresponds to the
propagation along the $c$-axis. We have used a ``fatband''
representation associated with the $\{A_{1g},E_g\}$ orbitals on
Fig.~\ref{fig:fatbands} in which the width is proportional to the amount of
orbital character of each band at a given $k$-point. Thus, one can
identify that the narrow bands at the Fermi level are associated with the
$E_g$ orbitals. Along $\Gamma$-Z starting at around -1 eV, one observes the
dispersive band with strong $A_{1g}$ character that crosses $\varepsilon_F$
close to the edge of the BZ. Hence the $\Gamma$-Z portion of that band is
almost filled, with $2k_F$$\simeq$$0.94c^*$ as mentioned above. The $E_{g2}$
electron pocket at the $\Gamma$-point is absent for the hexagonal high-$T$
structure (see also \cite{mattheiss_bavs3_1995}). Its existence might be
related to the hole-like transport below $T_S$ as revealed from Hall
measurements~\cite{booth_bavs3_field_prb_1999}.
The partial density of states (DOS) for each orbital displayed on
Fig.~\ref{fig:vdos}, show a rather broad $A_{1g}$ DOS while the $E_g$'s yield
very narrow peaks right at and above $E_F$.
%
%

\begin{figure}
\epsfclipon
\epsfig{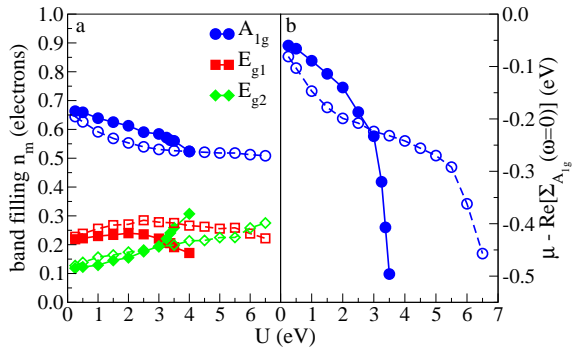}
\caption{(a) Band fillings at $\beta$=$(k_{\sub{B}}T)^{-1}$=15 eV$^{-1}$
         ($T$=$774$~K) for the effective bands within LDA+DMFT. (b)
         Corresponding shift of the Fermi level for the $A_{1g}$ band (note that
         $\varepsilon_{\vk_F}^{\sub{(LDA)}}$=0). Filled symbols: $U/J$=7,
         open symbols: $U/J$=4.}
\label{fig:occup}
\end{figure}

\begin{figure} 
\epsfclipon
\epsfig{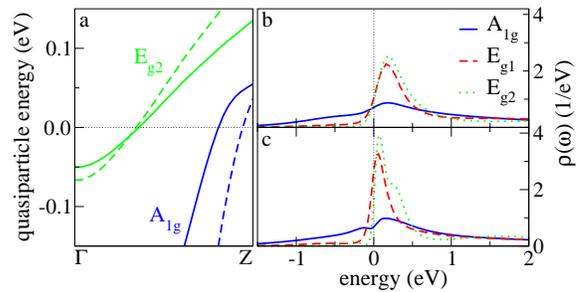}
\caption{LDA+DMFT spectral data for $U$$=$3.5 eV, $U$$/$$J$=4. (a) V($3d$)
         low-energy quasiparticle bands along $\Gamma$-Z in LDA (dashed lines)
         and LDA+DMFT (solid lines) for $T$=332~K. (b,c) integrated spectral
         function $\rho(\omega)$ for a single formula unit of BaVS$_3$ at
         $T$=1160~K (b) and $T$=332~K (c).}
\label{fig:specfunc}
\end{figure}

\begin{figure} 
\epsfclipon
\epsfig{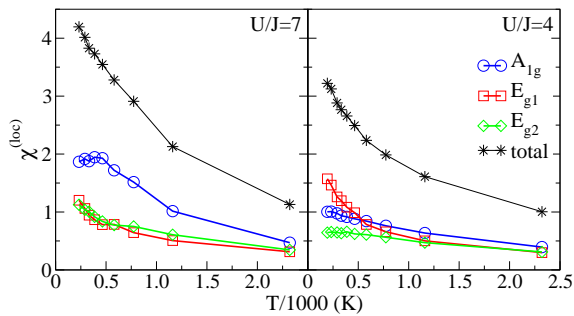}
\caption{$T$-dependent local spin susceptibilities
         for $U$=3.5 eV, according to the normalization
         $\chi^{\sub{(loc)}}$=$\int_0^\beta d\tau\langle
         \hat{S}_z(0)\hat{S}_z(\tau)\rangle$, where $\hat{S}_z$ denotes the
         $z$-component of the spin operator.}
\label{fig:chiloc}
\end{figure}

Recently, it has become possible to investigate correlation
effects in a realistic setting by combining LDA with 
DMFT~\cite{anisimov_lda+dmft_1997,lichtenstein_lda+dmft_1998}.
Starting from the LDA hamiltonian $H^{\rm{LDA}}_{mm'}(\vk)$
expressed in a localized basis set, many-body terms are
introduced, leading to a self-energy matrix $\Sigma_{mm'}$ which
is taken to be local (\vk-independent) but fully frequency-dependent. In the 
present work, we use a simplified implementation of the LDA$+$DMFT approach, 
which is geared at keeping those physical ingredients which are important for 
the physics of \bavs3 close to the MIT. 
Specifically, we work within an effective 3-band model, whereby
the LDA electronic structure enters via the DOS in the relevant energy window 
around the Fermi level. The effective bands
are derived from the symmetry-adapted $\{A_{1g},E_g\}$ states, thus
non-diagonal self-energy terms $\Sigma_{m\neq m'}$ are negligible.
A physically adequate `empirical downfolding' procedure was used in order to
construct the effective 3-band DOS. First, we took care of the 3$d^1$
character. The location and width ($\sim$2.2 eV) of the energy window
were chosen such that the total DOS accomodates two electrons below and ten
electrons above the Fermi level (per two V). Since the $e_g$ bands are hardly
relevant and of minor weight close to $E_F$, they were hybridized with
the $E_g$ bands as suggested from the resolved partial DOS on
Fig.~\ref{fig:vdos}. The contribution of these new $E_g$ bands was
substracted from the total DOS normalized to a single
formula unit of BaVS$_3$ within the chosen energy window. The resulting
difference was identified as the new downfolded $A_{1g}$ band, since the
$A_{1g}$ orbital substantially hybridizes with the S($3p$) orbitals.
Finally, the self-energies $\Sigma_m$ associated with these effective bands
are calculated from LDA$+$DMFT where the self-consistency condition is expressed 
as an integral over the effective partial DOS $D_m^{\rm{LDA}}(\varepsilon)$.
The on-site interaction matrix was parametrised as:
$U_{mm}^{\uparrow\downarrow}$=$U$,
$U_{m\neq m'}^{\uparrow\downarrow}$=$U$$-$$2J$ and
$U_{m\neq m'}^{\uparrow\uparrow(\downarrow\downarrow)}$=$U$$-$$3J$,
with $U$ the on-site Coulomb repulsion and $J$ the local Hund's rule coupling.
The DMFT local impurity problem was solved using the quantum Monte-Carlo (QMC)
Hirsch-Fye algorithm. Up to 128 slices in imaginary time $\tau$ and at most
$10^6$ sweeps were used~\footnote{The QMC calculations were performed
at temperatures significantly higher than the physical
\Tmit (cf. Figs.~\ref{fig:occup}-\ref{fig:chiloc}). However, we still think that
the essential physics close to \Tmit is captured. This is because the
coherence scale for the broad \a1g band is already reached for elevated
temperatures over most of the range of parameters that we have studied, and the
crystal structure at $T$=100 K was used in the LDA calculations.}.

%
%
Fig.~\ref{fig:occup}a displays our results for the occupancies of each orbital,
as a function of $U$.
In the absence of a precise determination of this parameter from
either experiments (e.g., photoemission) or theory (constrained LDA methods tend
to underestimate the screening for metals), we varied $U$ over a rather
large range of values. The ratio $U/J$ was chosen to be fixed, and two series
were studied: $U/J$=7 and $U/J$=4, corresponding to a weaker and stronger
Hund's coupling, respectively.
%
The orbital occupancies in our effective 3-band model, at the LDA level
(i.e., for $U$=$0$) read: $n(A_{1g})$=0.712, $n(E_{g1})$=0.207 and
$n(E_{g2})$=0.081. The main effect apparent on Fig.~\ref{fig:occup}a is that
moderate correlations tend to bring the occupancies of each orbital closer to
one another, i.e., to decrease the population of the ``extended'' orbital \a1g
and to increase the occupancy of each $E_g$ orbital. For strong correlations,
values close to $n(A_{1g})$$\simeq$$n(E_{g1})$$+$$n(E_{g2})$$\simeq$0.5
are obtained in the DMFT calculation, corresponding to a half-filled band.
In the absence of correlations, it pays to occupy dominantly the broad \a1g 
band, which provides the largest kinetic energy gain, while in the presence of 
correlations this has to be balanced versus the potential energy cost. The 
Hund's coupling clearly favors such an interorbital charge redistribution, as 
also pointed out recently in the context of ruthenates~\cite{okamoto_2004}.
For larger $U$, band-narrowing effects reduce the kinetic
energy of the \a1g band, hence also reducing its occupancy~\cite{lie00}
in favor of $E_g$.
As expected, a Mott insulating state is obtained when $U$ is increased beyond
a critical value, which is found to depend strongly on $J$ (the range of $U$'s
displayed in Fig.~\ref{fig:occup} corresponds to the metallic regime
for each series, on which we focus in this paper).
%

Our calculations reveal that the depletion of the \a1g band is accompanied by
a reduction of the corresponding $k_F$ along the $\Gamma$-$Z$
direction. (Note that the Luttinger theorem \cite{lut60} does not apply
separately for each band but only relates the total Fermi surface volume to the
total occupancy).
While a full determination of the quasiparticle (QP) band structure in the
interacting system requires a determination of the real-frequency
self-energy, we can extract the low-energy expansion of this
quantity from our QMC calculation, in the form:
$\mbox{Re}\Sigma_m(\omega+i0^+)$$\simeq$$
\mbox{Re}\Sigma_m(0)+\omega(1-1/Z_m)+\cdots$, with $Z_m$ the QP
residue associated with each orbital. The poles of the
Green's function determine the QP dispersion relation:
$\mbox{det}[\omega_{\vk}-\hat{Z}[\hat{H}_{\vk}^{\rm{LDA}}+\mbox{Re}\hat{\Sigma}
(0)-\mu]]$=0,
with $\mu$ the chemical potential.
Focusing first on the \a1g sheet of the Fermi surface, within our diagonal 
formulation the location of the Fermi wave vector in the interacting 
system is determined by:
$\varepsilon^{\rm{LDA}}_{A_{1g}}(\vk_F)$=$\mu-\mbox{Re}\Sigma_{A_{1g}}(0)$.
This quantity therefore yields the energy shift of the \a1g band at the
Fermi surface crossing, as compared to LDA. It is depicted in
Fig.~\ref{fig:occup}b as a function of $U$.
On Fig.~\ref{fig:specfunc}a, we display (for $U$=3.5eV and $U/J$=4) the QP 
bands that cross the Fermi level along $\Gamma$-Z in a narrow energy range 
around $\varepsilon_F$. The QP bands are obtained by performing a perturbative 
expansion of the pole equation above, which yields:
$\omega_{b\vk}$=$\sum_m C_{m\vk}^{b}Z_m[\varepsilon_{b\vk}^{\rm{LDA}}
+\mbox{Re}\Sigma_m(0)-\mu]$ with
$C_{m\vk}^{b}$$\equiv$$|\langle\psi_{\vk}^b|\phi_m\rangle|^2$
the LDA orbital weight. From these two figures, it is clear that $k_F(A_{1g})$ 
is reduced in comparison to the LDA value, in line with the global charge 
transfer from \a1g to $E_g$. This opens new possibilities for the CDW instability,
in particular for the nesting condition.

The enhanced population of the narrow $E_g$ bands, as
well as the correlation-induced reduction of its bandwidth
(see Fig.~\ref{fig:specfunc}a) provide an explanation
for the local moments observed in the metallic phase.
To support this, we have calculated (Fig.~\ref{fig:chiloc})
the local susceptibility associated with each orbital
$\chi^{\rm{(loc)}}_m\equiv\sum_{{\bf q}}$Re$[\chi_m({\bf q},\omega$=0)].
For both values of $U/J$, the susceptibility of the \a1g band
saturates to a Pauli-like value at low temperatures. In contrast, 
$\chi^{\rm{(loc)}}$ of the $E_g$ orbitals strongly increases as $T$
is lowered (except for the low-filled $E_{g2}$ orbital at $U/J$=4).
This is because the coherence temperature below which quasiparticles
form is much lower for the $E_g$ orbitals than for the \a1g.
Accordingly, our calculation of the integrated spectral functions
(Fig.~\ref{fig:specfunc}b,c) reveals a strong T-dependence of the $E_g$
QP peak. Some differences between the two series are clear from
Fig.~\ref{fig:chiloc}. For $U/J$=7, the system is very close to the Mott 
transition. Thus the \a1g electrons also act as local moments over part of the 
temperature range, while for $U/J$=4, the $T$-dependence of the total
local susceptibility is almost entirely due to the $E_{g1}$ electrons.
Which of the two situations is closest to the physics of \bavs3
does require further investigations, albeit
some experimental indications point at the second possibility~\cite{Faz01}.

In conclusion, we have shown that charge redistributions due to correlations
lower the occupancy of the broad \a1g band in favor of the narrow $E_g$'s, in
comparison to LDA calculations. This explains the presence of local moments in 
the metallic phase and paves the road towards a full understanding of the CDW 
instability of this material. Orbital-selective experimental probes are highly 
desirable in order to put to a test the conclusions of the present work. Among 
several outstanding questions still open are the detailed nature of the 
insulating phase (in particular regarding the partial suppression of local 
moments and the eventual magnetic ordering) as well as the remarkable suppression
of \Tmit under
pressure~\cite{graf_bavs3_pressure_prb_1995,forro_bavs3_qcp_prl_2000}.
\acknowledgements
We are grateful to J.-P.~Pouget and L.~Forr\'{o} for stimulating our interest
in this material, and acknowledge many useful discussions with the Orsay
experimental group (S.~Fagot, P.~Foury-Leylekian, J.-P.~Pouget and
S.~Ravy), and
also with P.A.~Lee and T.~Giamarchi. Computations were performed at IDRIS
Orsay.\\
\textsl{Note added}. After the completion of this work, we became aware of a
recent angle-resolved photoemission study of BaVS$_3$ by S.~Mitrovic
\textsl{et al.} supporting our prediction of a reduced $k_F(A_{1g})$.
\bibliographystyle{apsrev}
\bibliography{bibag,bibextra}

\end{document}